\documentclass[3p,times,twocolumn]{elsarticle}

\usepackage{ecrc}
\usepackage{aas_macros}
\volume{00}
\firstpage{1}

\usepackage{lineno,hyperref}

\journalname{Journal of High Energy Astrophysics}
\journal{Journal of High Energy Astrophysics}

\def\nup{$\nu^{S}_{\rm peak}$}

\def\gr{$\gamma$-ray}
\def\nufnu{$\nu f(\nu)$}

\runauth{P. Giommi}
\jid{nuphbp}

\jnltitlelogo{Journal of High Energy Astrophysics}

\begin{document}
\begin{frontmatter}

\title{Multi-frequency, multi-messenger astrophysics with Swift. \\ The case of blazars}

\author{Paolo Giommi\fnref{paolo.giommi@asdc.asi.it}}
\address[mymainaddress]{ASDC, Agenzia Spaziale Italiana, Via del Politecnico s.n.c., I-00133 Roma, Italy,}
\address[mysecondaryaddress]{ICRANet-Rio, CBPF, Rua Dr. Xavier Sigaud 150, 22290-180 Rio de Janeiro, Brazil}
\fntext[myfootnote]{paolo.giommi@asdc.asi.it}

\begin{abstract}

During its first 10 years of orbital operations Swift dedicated approximately 11\% of its observing time to blazars, carrying out more than 12,000  observations of $\sim$1,600 different objects, for a total exposure time of over 25 million seconds.  
In this paper I briefly discuss the impact that Swift is having on blazar multi-frequency and time-domain astrophysics, as well as how it is contributing to the opening of the era of multi-messenger astronomy.
Finally, I present some preliminary results from a systematic analysis of a very large number of Swift XRT observations of blazars. All the "science ready" data products 
that are being generated by this project will be publicly released. Specifically, deconvolved X-ray spectra and best fit spectral parameters will be available through the 
ASDC "SED builder" tool (https://tools.asdc.asi.it/SED) and by means of interactive tables (http://www.asdc.asi.it/xrtspectra).  Innovative data visualisation methods 
(see e.g. http://youtu.be/nAZYcXcUGW8) are also being developed at ASDC to better exploit this remarkable and rapidly growing data set.

\end{abstract}

\end{frontmatter}


\section{Introduction}

Blazars are a rare type of AGN that are well known for their extreme observational properties such as superluminal motion and highly variable 
non-thermal emission across the entire electromagnetic spectrum, from radio waves to the highest energy \gr s. 
Their unique properties are attributed to emission of radiation by energetic particles that move towards us in a magnetic field within a relativistic jet that 
happens to be pointing close to our line of sight \citep{bla78,urry95}.  

\begin{figure}
\includegraphics[width=1.0\linewidth]{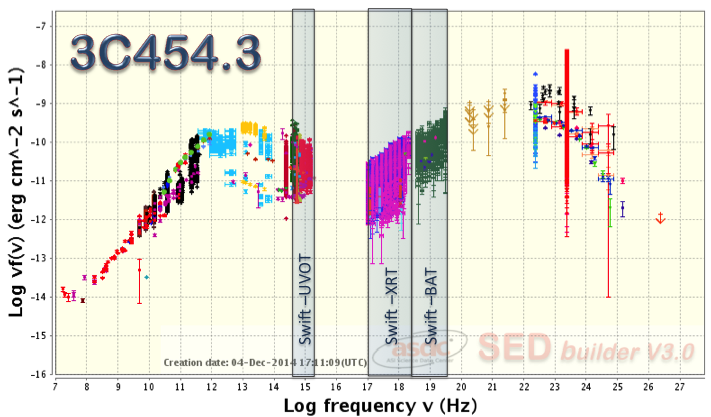}
\caption{The SED of the blazar 3C454.3 assembled with over 60,000 flux measurements obtained with many space and ground-based observatories. The contribution of 
all Swift instruments (UVOT, XRT and BAT) is highlighted by the three grey vertical bands.}
\label{fig:3c454p3}
\end{figure}

Blazars come in two flavours: Flat Spectrum Radio Quasars (FSRQs ) and BL Lacertae objects (BL Lacs). The main difference is in their optical spectra, with the former displaying broad emission lines 
and the latter being instead characterised by either a featureless continuum, or by a spectrum that displays only absorption features (usually from the host galaxy) or weak
narrow emission lines \cite{giommi2012}.

The radio to $\gamma$-ray Spectral Energy Distribution (SED) of blazars typically displays two broad humps when plotted in \nufnu\ vs $\nu$ space (see e.g. fig. \ref{fig:3c454p3}).

It is widely accepted that the radiation associated to the low-frequency hump in the SED of blazars is due to synchrotron emission from relativistic electrons moving in a magnetic field within the jet.
The nature of the high energy emission, may instead be attributed to two intrinsically different mechanisms, described within scenarios generally referred to as leptonic and hadronic models \cite{boettcher2013}. 
In leptonic scenarios the $\gamma$-ray emission is assumed to be due to inverse Compton radiation \cite{maraschi92,sikora94} whereas in leptohadronic models it
could be due to proton synchrotron radiation \cite{aharonian00,mucke01} or it may have a photohadronic origin \cite{petropoulou15}

The peak energy of the synchrotron component (\nup) ranges from about 10$^{12.5}$ Hz (equivalent to approximately 0.01~eV) to over 10$^{18.5}$  Hz ($\sim$13 KeV), reflecting the maximum 
energy at which particles can be accelerated \cite{giommi2012}.  
Sources where 
\nup is less than 10$^{14}$ Hz ($\sim$0.4 eV) are called low synchrotron peaked blazars (LBL), while those where \nup\ $>$ 10$^{15}$ Hz ($\sim$4 eV)
are called  high energy peaked blazars, or HBL \citep{padgio95}.

For a long time blazar research has been confined to a small community of specialists. This is rapidly changing as these sources are receiving much more general attention after they have been found to be the largest population of extragalactic sources in microwave (between $\sim$ 30 and $\sim$ 200 GHz), and in $\gamma$-ray (0.1$<$ E $<$ 100 GeV) surveys, as well as the most common extragalactic objects appearing in catalogs of TeV detected sources. Most of the blazars found in microwave surveys are of the LBL type (both FSRQs and BL Lacs), while HBL BL Lacs dominate the \gr\ and the VHE (E $>$ 100 GeV) extragalactic sky.

Blazars are powerful extragalactic sources capable of accelerating particles to very high energies and therefore they are considered as prime candidates for multi-messenger astrophysics.


\section{Swift and blazars}

\begin{figure}
\includegraphics[width=1.0\linewidth]{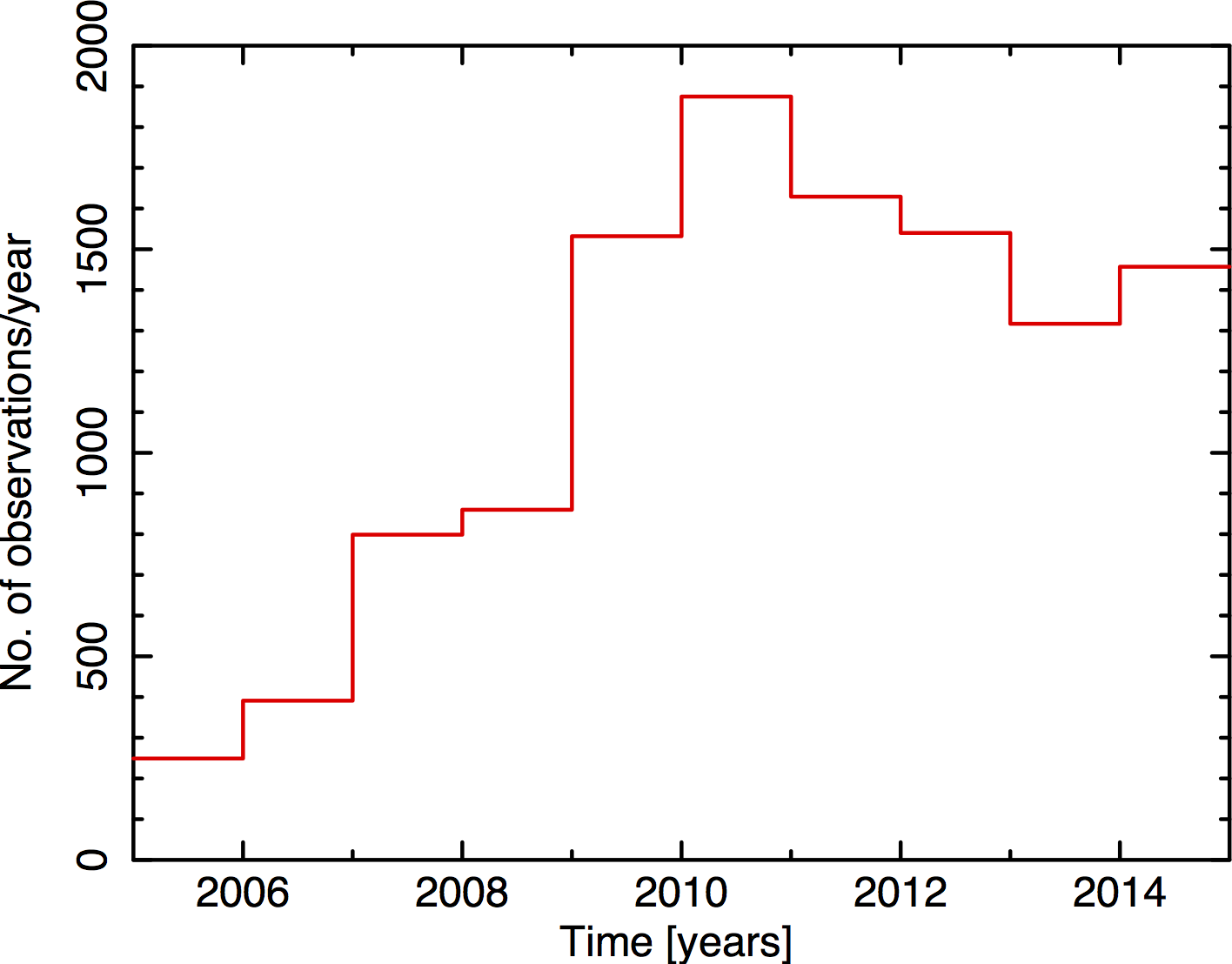}
\caption{The rate of Swift observations including at least one blazar in the XRT field of view. 
Note the rapid increase from the beginning of the scientific operations in early 2005 up to 2010, after which date the rate decreased somewhat stabilising 
around an average level of $\approx$1500 observations/year.}
\label{fig:obsprofile}
\end{figure}

During its first ten years in orbit the Swift observatory \cite{swiftpaper} carried out more than 12,000 observations of approximately 1,600 blazars, for a total  
of $\approx$ 25.3 Ms of X-Ray Telescope (XRT, \cite{burrowsXRT}) net exposure time, or approximately 11\% of the entire scientific program. About 50\% of the objects listed in the most up to date catalog of blazars (BZCAT, 5$^{\rm th}$ edition \cite{5bzcat}) have been observed at least once. Some of the well known and best studied objects have been observed hundreds of times.  Examples are the BL Lacs MRK 421 with 635 observations, MRK501 with  412 observations, and the FSRQ 3C454.3 with  419 pointings.
Fig. \ref{fig:obsprofile} illustrates how the rate of Swift blazar observations evolved in time, showing a rapid increase between 2005 and 2009, followed by an approximately constant level of 1,500 observations per year.  This is quite a large number of pointings reflecting a scientific strategy which reserves to blazars a sizeable fraction of the Swift overall observing program.
 The full data set is available from the official Swift archives in the U.S. (http://swift.gsfc.nasa.gov/), Italy (http://swift.asdc.asi.it/), and U.K. (http://www.swift.ac.uk/). 
 However, current archive facilities only provide basic data sets and software packages for the data reduction. Scientific analysis must be carried out by the user by means of mission specific and additional analysis software. This is usually done for  a single exposure or for a limited number of observations; clearly normal users could hardly manage
to analyse the entire data set. To facilitate the use of this valuable public archive at ASDC, also as part of the Swift team, we started a program to systematically analyse Swift blazar data to provide higher level  "science ready" products and results, such as deconvolved spectra and best fit parameters for different spectral models.

\begin{figure*}
\includegraphics[width=1.0\linewidth]{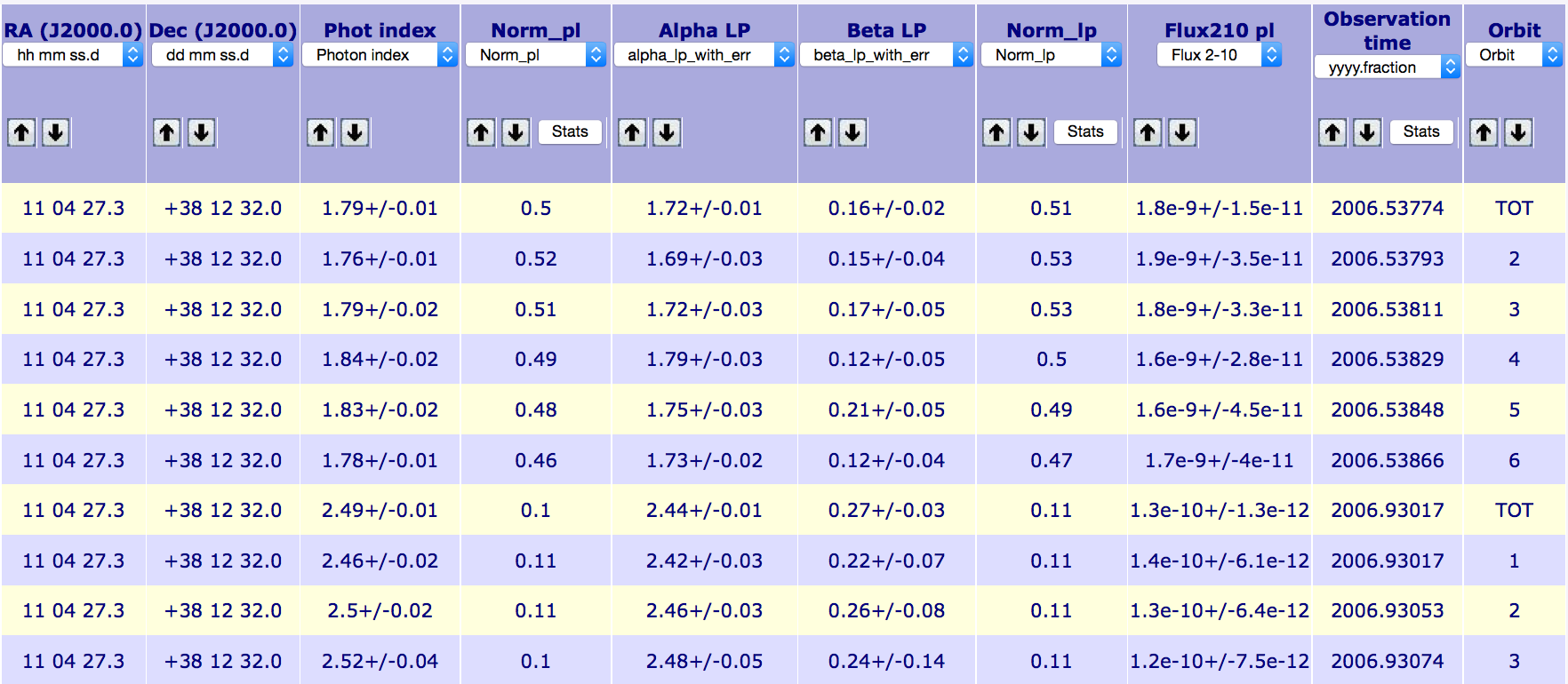}
\caption{The first few lines of the interactive table through which the results of the ASDC XRT spectral analysis of blazars will be available on-line at http://www.asdc.asi.it/xrtspectra/}
\label{fig:xrttable}
\end{figure*}

An example of the remarkable contribution that Swift is providing to multi-frequency studies of blazars is given by the SED of 3C454.3 shown in 
Fig. \ref{fig:3c454p3} where the simultaneous data obtained from the three Swift instruments, UVOT, XRT and BAT, during the 419 observations carried out before the end of 2014, 
is highlighted by three vertical bands.

\section{Swift XRT Data Analysis}

At the ASI Science Data Center (ASDC) we have processed in a uniform way, using the latest reduction software and calibration available,  all  XRT observations of 
blazars pointed by Swift at least 10 times (approximately 120 blazars). 
The processing was carried out both on the overall data collected during each observation and also on a orbit by orbit basis, thus allowing a time resolution of the order
of 1.5 hours. 

\begin{figure}
\includegraphics[width=1.0\linewidth]{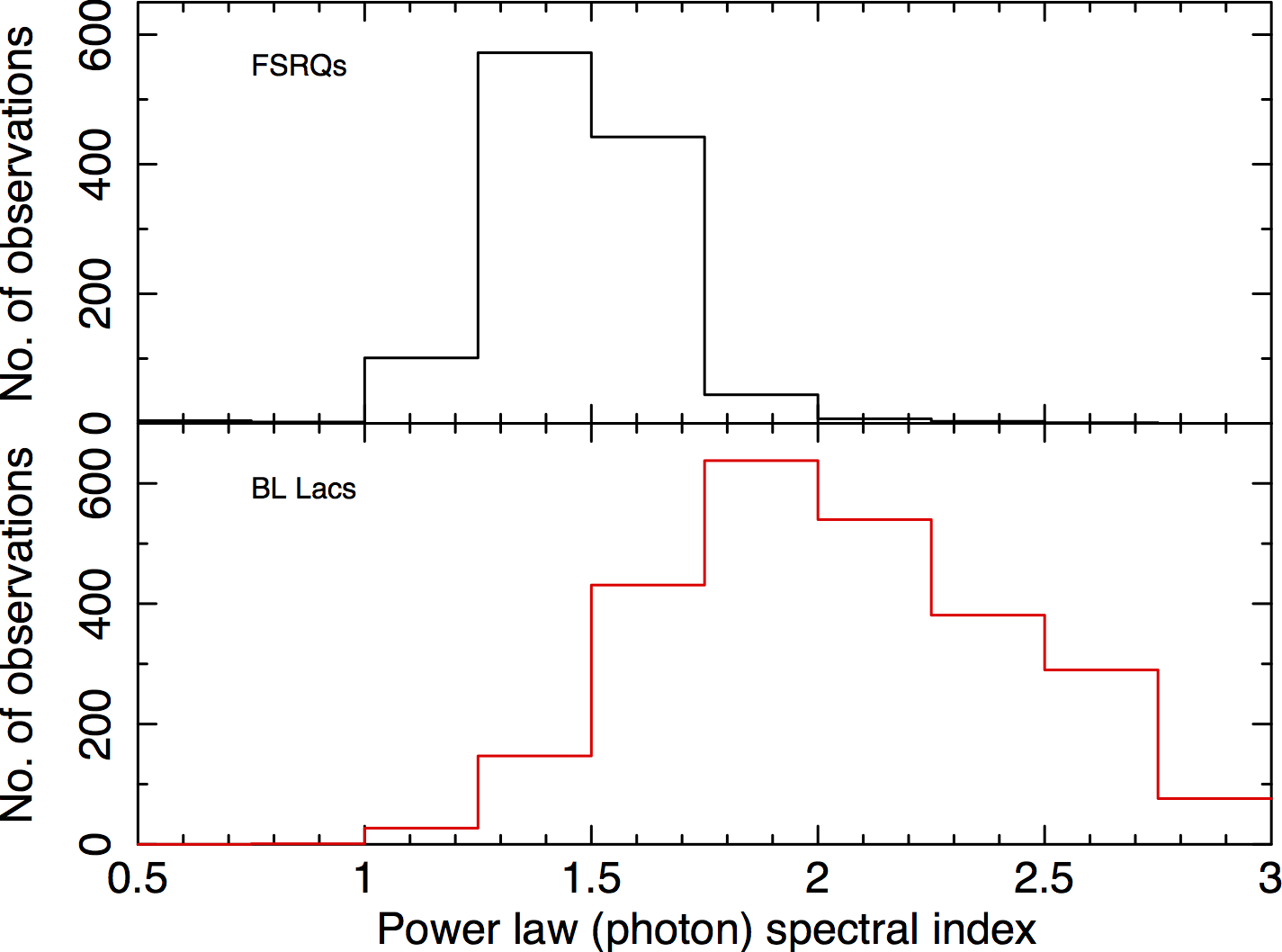}
\caption{The distribution of spectral slopes of the blazars most observed by Swift. FSRQs  are shown in the upper panel, BL Lacs in the lower panel.}
\label{fig:spectral_slopes}
\end{figure} 

The results of  the analysis of approximately 5,000 XRT observations (and nearly 9,000 orbits) will be published shortly \citep{giommi2015a}. The high level data products, such as 
deconvolved spectral data, light-curves, best fit parameters etc, will be available on-line within the ASDC SED builder tool (https://tools.asdc.asi.it/SED) and through interactive tables  (http://www.asdc.asi.it/xrtspectra),  a preview of which is shown in fig. \ref{fig:xrttable}. Best fit spectral parameters are given for the most commonly used spectral models (power law and log parabola), for each observation and for every orbit during which sufficient statistics is collected for a spectral analysis.  UVOT and BAT data for the same blazar sample will 
also be processed and the  results will be published in a similar way in the future. In the following we show some preliminary results of this work, which will lead to the establishment of one of the largest existing databases of high level scientific products and results on blazars. 
 
\begin{figure}
\includegraphics[width=1.0\linewidth]{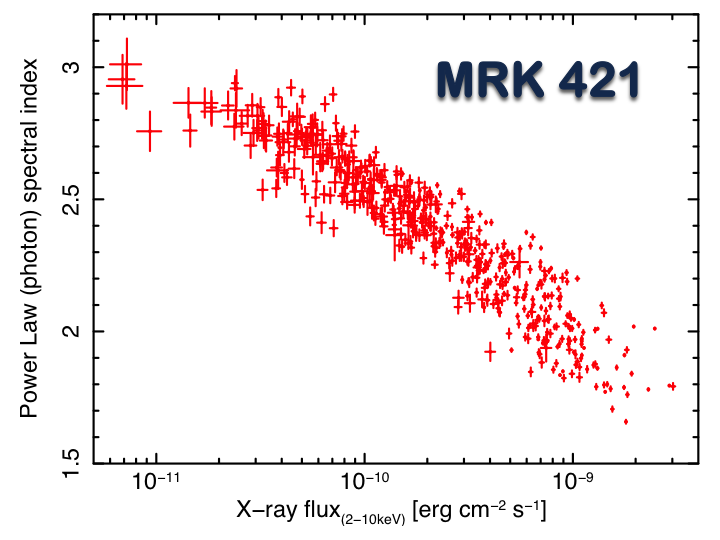}
\caption{The correlation between X-ray spectral slope and X-ray flux in the BL Lac object MKN421. Note the very large range in flux (a factor of $\approx$ 500) and spectral 
slope. More than 600 Swift observations are shown in this plot.}
\label{fig:mkn421f}
\end{figure}

One example of these results is illustrated in fig \ref{fig:spectral_slopes} where the distributions of the best fit power law spectral indices are shown for the subsamples of FSRQs and BL Lacs. Fig. \ref{fig:mkn421f} shows instead how the spectral slope is related to source intensity for the case of MRK 421. The tight correlation apparent from this plot 
 is well known \cite[e.g.][]{giommi90,tramacere09}, however the very large amount of data plotted in fig. \ref{fig:mkn421f} provides unprecedented statistics both in spectral slope and in flux range, this last spanning a factor of $\approx$500, probably a record value for soft X-ray variability of AGN.  
  
\section{Time domain data}

The remarkable number of measurements accumulated over the last several years, thanks to Swift and many other space and ground-based observatories, are enabling new approaches to the analysis of luminosity variability in all parts of the electromagnetic spectrum. In this contribution blazars are used to provide some examples of how this new 
opportunity is being exploited.
\begin{figure}
\includegraphics[width=1.0\linewidth]{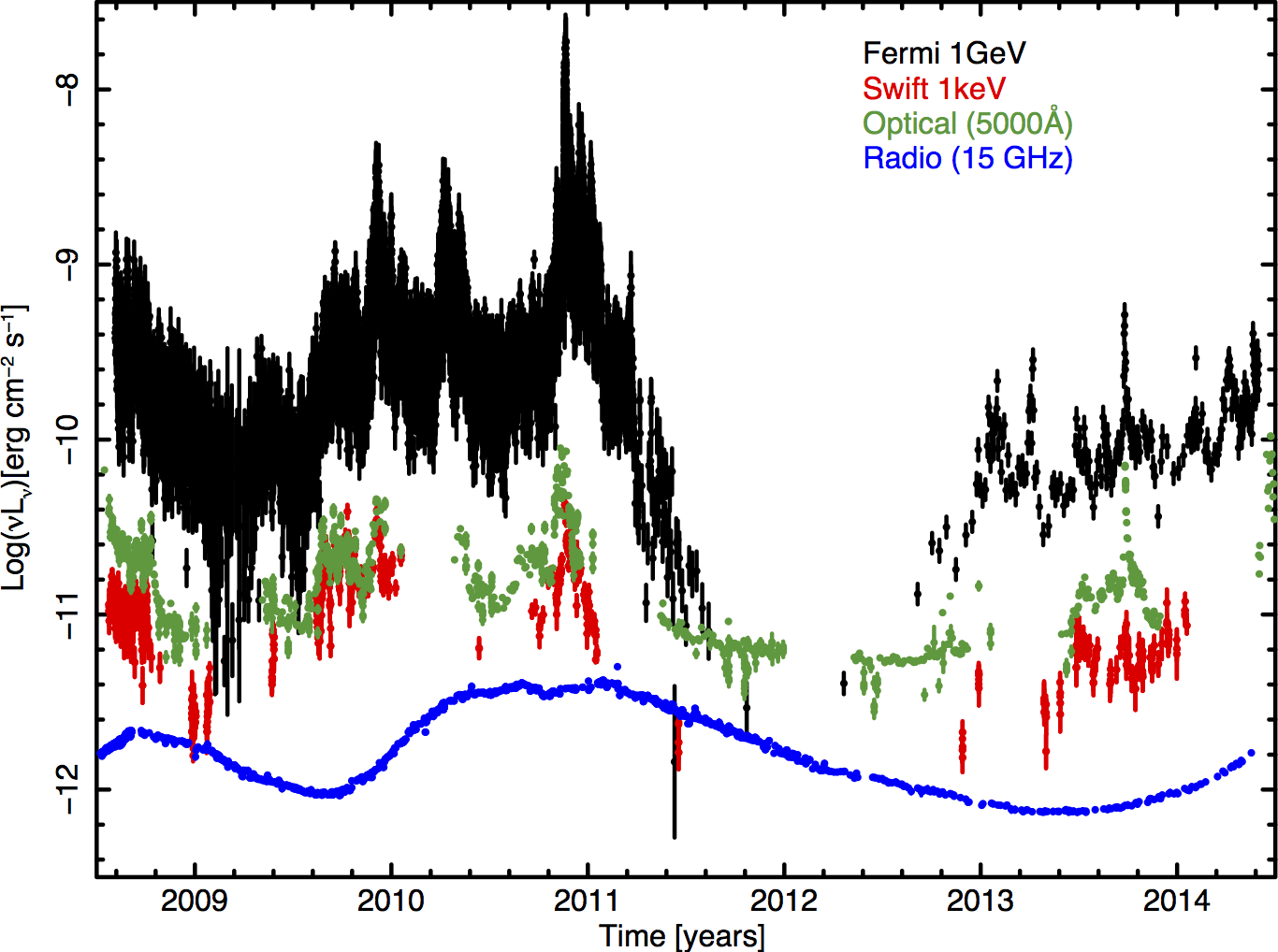}
\caption{Multi-frequency light-curve of the blazar 3C454.3. In this plot the same \gr, X-ray, optical and radio data extracted from the data set used for the 
SED of fig\ref{fig:3c454p3} is used to illustrate the different behaviour in this source at different energies.}
\label{fig:mflc}
\end{figure}

\begin{figure}
\includegraphics[width=1.0\linewidth]{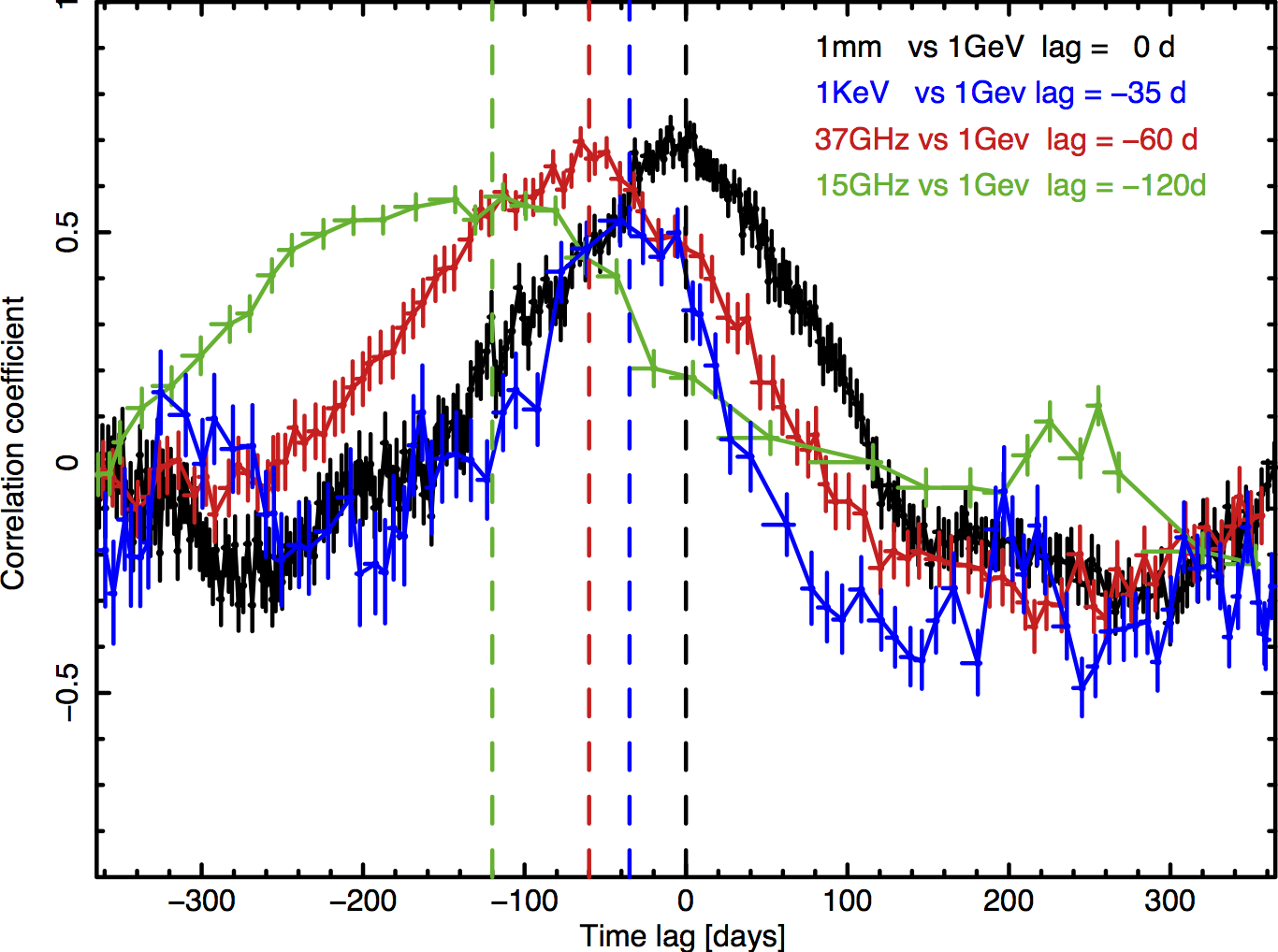}
\caption{The Discrete Cross Correlation Function on multi-frequency data of 3C345.3 illustrating how different energy bands (15 GHZ, 37 GHz, 1mm, and 1 keV) show different time
delay with respect to the \gr\ emission at 1 GeV.}
\label{fig:dcf}
\end{figure}

Fig. \ref{fig:mflc} shows the light-curve of 3C454.3 in the \gr, X-ray, optical and radio band. It has been built using the same data set used to assemble the SED of 
fig. \ref{fig:3c454p3}. 
A simple visual inspection of this plot reveals that the light-curves at different frequencies share a similar overall behaviour, but also show large differences.
One technique oftenÊused to quantify these differences is the discrete cross-correlation function (DCCF) \cite{alexander} between the light-curves at different energies.
Fig. \ref{fig:dcf} shows the DCCF of 3C454.3 obtained by comparing the \gr\ light-curve (1~GeV from Fermi-LAT observations) with data taken in different energy bands, namely 
1~keV, 1 mm,  37 GHz and 15 GHz. As can easily be seen in fig. \ref{fig:dcf} the amount of correlation with \gr s is maximum for 1 mm data,
with no significant time lag. All other light-curves compared to 1GeV data show a lower lever of correlation and significant time lags, ranging from about one month 
at 1KeV to several months in the radio band, demonstrating the complexity of the emission processes in blazars, which are subject to different dynamical timescales in 
various parts of the electromagnetic spectrum. 
This implies that while the use of simultaneous data, often obtained through complex observational campaigns involving manyÊground and space-based observatories (see e.g. \cite{giommiplanck}), is 
desirable and probably necessary, it is not sufficient for a  full understanding of the physics of blazars. 

\begin{figure}
\includegraphics[width=1.0\linewidth]{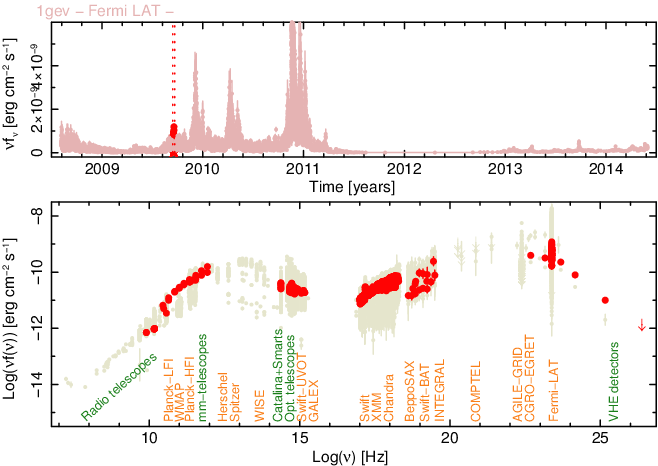}
\includegraphics[width=1.0\linewidth]{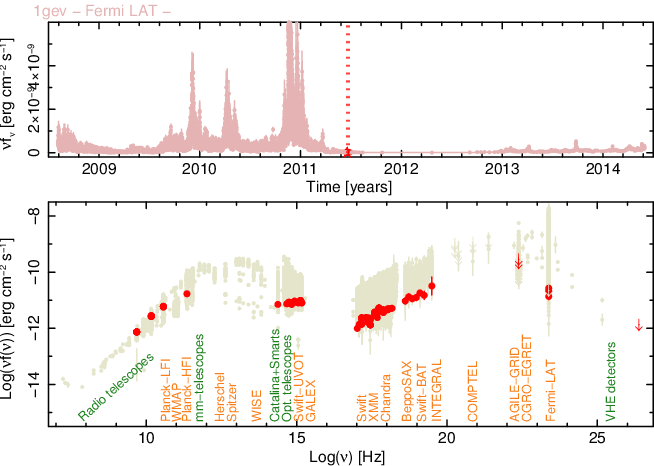}
\caption{Two frames extracted from the SED movie of the blazar 3C454.3 that can be watched at http://youtu.be/nAZYcXcUGW8. 
The red points are simultaneous multi-frequency measurements taken within the time interval marked by the red dashed lines 
in the top part of each frame showing the 1GeV light-curve.  Green coloured text represents ground-based telescopes, orange text indicates satellite observatories.
The top frame shows the source in a state of growing intensity in which the emission is very large at all frequencies, with the exception of the radio and microwave bands. The 
lower frame illustrates an intensity state where the source is in a decreasing phase of its overall activity, but it is still bright at radio and microwave frequencies.}
\label{fig:3c454movie}
\end{figure}

A completely new technique to visualise variability in the SED of cosmic sources is to run in a sequence a set of frames, each representing the SED in a particular time interval, 
just like in a movie. Of course, this can be done only if a sufficiently large number of multi-frequency measurements are available at all times. This is a requirement that is quite 
demanding but that has been already met by some of the bright and well known sources. As an example, fig. \ref{fig:3c454movie} shows two frames of such a movie, which was built combining over 60,000 multi-frequency flux measurements of the blazar 3C454.3. The full movie can be accessed on-line at http://youtu.be/nAZYcXcUGW8 .

\section{Blazars and multi-messenger astrophysics}

Due to their large luminosity and SEDs often reaching the highest observed \gr\ energies, blazars are the most powerful and most energetic cosmic accelerators known. 

An estimate of the maximum energy at which particles are accelerated within this scenario can be obtained by measuring the peak (in \nufnu\ space) of the synchrotron hump, 
\nup. Swift is sensitive in the soft and hard X-ray bands, where the \nup\ of the most energetic blazars is located. Having observed about 50\% of the known blazars the 
Swift archive provides the largest potential for the measurement of the synchrotron peak, and therefore the maximum energy at which particles can be accelerated in blazars. 
 
\subsection{Neutrinos from blazars?}

Since their detection at very high $\gamma$-ray energies by Cherenkov telescopes, HBL blazars (which are the type of blazar most frequently detected in the VHE band) have been considered as 
candidate sources of extragalactic neutrinos \cite{halzen,mannheim99}.

The accuracy of the arrival direction of neutrinos as measured by current detectors, however,  can be as large as several degrees, so a pure positional matching is hardly sufficient to guarantee a statistically sound association.
Padovani \& Resconi (2014) \cite{Padresc} suggested that seven BL Lac objects of the HBL type could be related to as many neutrino events recently reported by the IceCube collaboration \citep{icecube} 
on the basis of both positional and energetic diagnostics. That is, not only candidate blazar counterparts are required to be within the positional uncertainly region of the IceCube neutrinos, but also their \gr\ intensity, which is assumed to be tightly connected to neutrino emission, must be comparable to the observed neutrino flux. 

\begin{figure}
\includegraphics[width=1.0\linewidth]{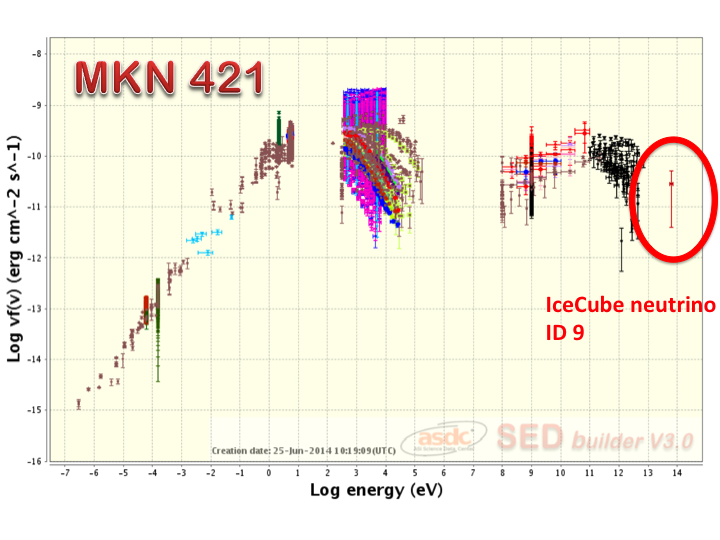}
\caption{The SED of MKN 421 plotted together with the flux expected from the twin-$\gamma$-ray photons associated to IceCube neutrino with id 9, assuming that the association reported in \cite{Padresc} is correct.}
\label{fig:mkn421}
\end{figure}

Fig. \ref{fig:mkn421} shows the SED of the BL Lac object MRK 421, (one of the sources reported in \cite{Padresc}) including the expected flux from the twin $\gamma$-ray photons  assuming that the IceCube neutrino with ID 9 is indeed associated to this bright BL Lac object. Petropoulou and collaborators \cite{petropoulou} recently showed that the SED of this object can be interpreted in terms of a one-zone lepto-hadronic model.
Assuming that this model is correct Padovani et al. 2015 \cite{padovani15}, have estimated the total neutrino cosmic background from the entire population of blazars and compared it with IceCube data.  
The expected overall extragalactic intensity is close to the observed one at energies above $\sim$ 1PeV, supporting the hypothesis that blazars as a class are a major producer of extragalactic neutrinos.

Other authors \cite{tanami} have also argued in favour of a possible association of radio-brightÊblazars (mostly of the LBL type) in the TANAMI sample with IceCube neutrinos. Also in this case this hypothesis is based on theoretical predictions rather than on a significant statistical signal. 

Upcoming deeper IceCube observations are clearly necessary to confirm or disprove the blazar-neutrino association.

\subsection{Could blazars be the sources of UHECRs?}

The origin of Ultra High Energy Cosmic Rays (UHECRs) is a major question that after many years of observations still remains unanswered \citep{abbasi2014}. Many 
such particles with energy exceeding 50 EeV have been detected by the most sensitive cosmic ray observatories such as the Telescope Array in the northern hemisphere 
and the Pierre Auger observatory in the south.  Their arriving directions are not correlated with the Galactic coordinates
and their energy is so large that they are not expected to be deflected much by magnetic fields therefore retaining the information about the source of origin.
High expectations have therefore been expressed in the literature for the association of UHECRs with extragalactic objects.
Attempts have been made to associate AGN to UHECRs \cite{george,agnauger2008} by cross-matching the positions of nearby AGN in the Veron-Cetty \& Veron (2006) catalog \cite{veron06} and 
of AGN detected in the Swift-BAT surveyÊwith Auger UHECRs finding what it seemed to be a likely association supported by an  encouragingly high confidence level. The search was limited to 
sources closer than $\sim$100 Mpc due to the fact that extragalactic UHECRs are expected to be severely attenuated by pion photoproduction interactions with the cosmic microwave background (CMB) 
radiation, the so called GZK effect.
This high level of statistical confidence, however, was not confirmed by subsequent searches carried out when larger samples of UHECRs became available \cite{macolino,aab15}. 
The disappointing outcome curtailed the high expectations for the discovery of the astrophysical counterparts of UHECRs. In fact, these initial expectations that might have been resting on optimistic confidence 
levels possibly estimated without fully taking into account of the {\it penalty } associated with the many attempts performed with different cuts and on different samples in the search for the long-awaited correlation with astrophysical sources. 

\begin{figure}
\includegraphics[width=1.0\linewidth]{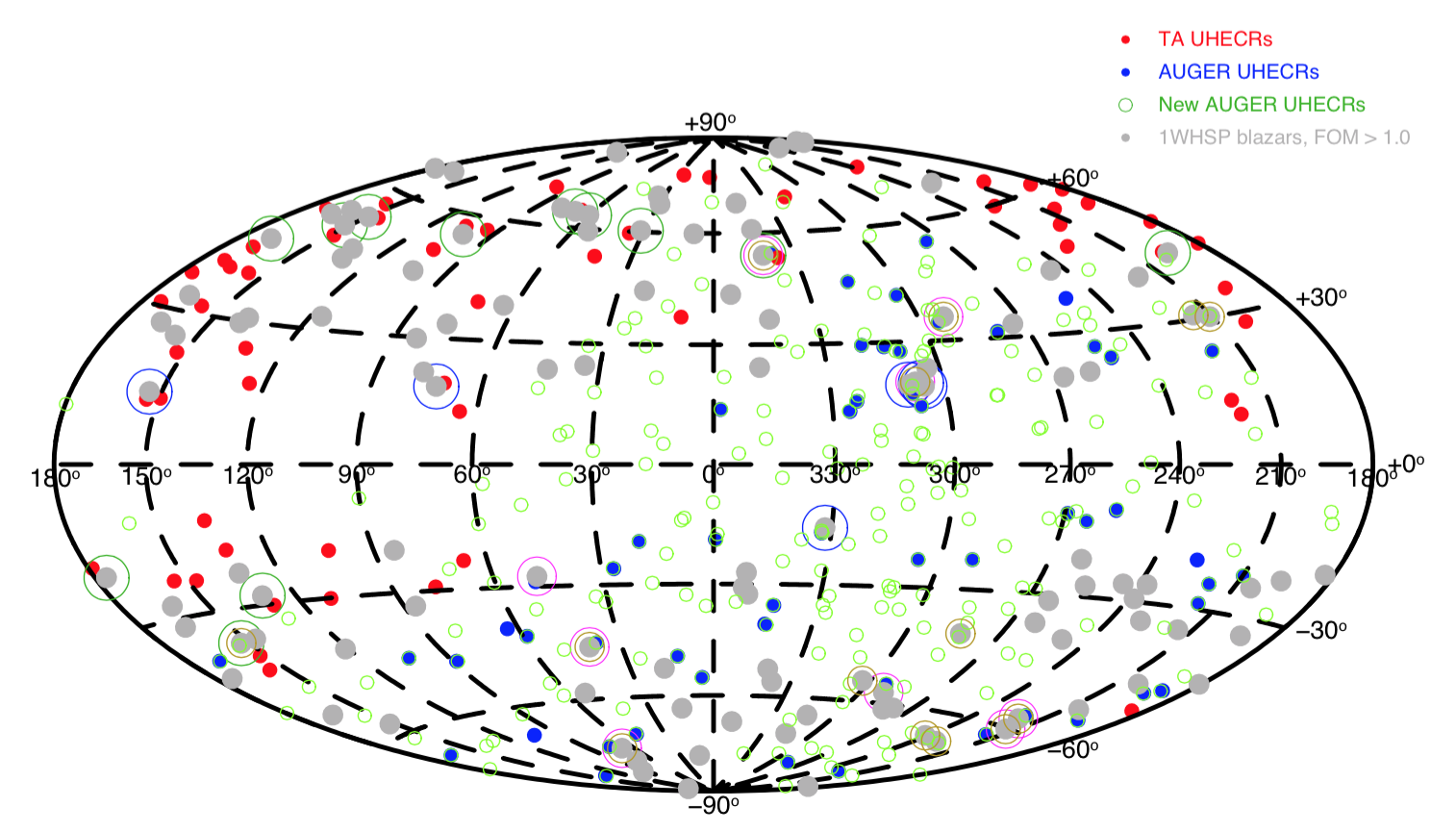}
\caption{Aitoff projection in Galactic coordinates of the UHECRs detected by the TA and Auger cosmic ray observatories, together with the positions of 1WHSP blazars that
could be detected by the current generation of Cherenkov telescopes. Large circles highlight the matches between UHECR and blazars.}
\label{fig:uhecr}
\end{figure}

These searches were however limited to mostly radio quiet AGN, that is sources that do not radiate at very high energy.
Blazars, particularly those with high energy synchrotron peak (HBL) that are often detected at VHE energies, are the extragalactic sources that are known to accelerate particles to the highest observed energies, and therefore are natural candidates as sources of UHECRs. In this context Tinyakov \& Tkachev (2001, \cite{tinyakov}) searched for and found a high confidence correlation between BL Lacs and UHECRs.  At ASDC in the months before the Swift 10 year meeting, we have tested this hypothesis by comparing the arriving directions of UHECRs from the TA and Auger samples with the position of the 1WHSP blazar sample \citep{arsioli2015}, the most complete and largest sample of blazars of the type that are expected to emit at VHE energies. In particular we used the subset of 110 1WHSP blazars that are bright enough to be detected by the current generation of Cherenkov telescopes, and cross-matched it with the arrival directions of TA and Auger UHECRs obtaining a combined probability of chance 
occurrence lower than 10$^{-3}$ \cite{arsiolithesis}. However, a preliminary verification of this encouraging result on the enlarged sample of UHECRs presented by the Auger collaboration in late 2014 \cite{aab} once again did  not confirm the significance of the correlation. To illustrate the status of the search at the time of writing fig. \ref{fig:uhecr} plots the arrival directions of the UHECRs from TA (red points)  and Auger (blue and green circles) in Galactic coordinates, together with the positions of the 1WHSP sources that are expected to be above the sensitivity threshold of the current generation of TeV telescopes (grey points).  The locations where there is a matching between blazars and UHECRs (within 3.2 degrees for the case of TA  and 2.5 degrees for Auger) are indicated by open large circles.  
  
Work is still in progress as we are currently using new Swift data to identify the sources with the largest synchrotron peak energy and therefore 
refine the sample by selecting the closest, most powerful and most energetic objects. The results will be presented in a future publication.

\section{Conclusion}

Since the beginning of its orbital operations Swift played a major role in accumulating what has now become one of the largest public archives of multi-wavelength data covering a time span of 
over ten years and including more than 12,000 observations of 1,600 blazars. This is certainly one of the richest existing multi-frequency open-access archive that is easily accessible through on-line 
services. 

The large potential for new discoveries and what can be done with this powerful archive has been illustrated with a number of examples.

In this paper I have also shown how, thanks to Swift and to many other space and ground-based astronomical facilities, innovative techniques are emerging enabling significant progress in the analysis 
of multi-frequency, multi-temporal astronomical data. These new techniques exploit the rich content of the available datasets in a deeper and more complete way compared to previous approaches, 
and will no doubt lead to discoveries that would not possible through a conventional approach, thus making archival research an increasingly effective tool for modern astrophysics.   

Multi-messenger astrophysics is the additional frontier. Although an irrefutable proof that blazars are connected to neutrinos or UHECRs has yet to be found, these peculiar objects 
firmly remain among the most promising extragalactic sources holding the potential for opening soon this new and exciting window on the Universe. 

\section {Acknowledgments}

Many of the results described in this paper have been obtained in cooperation with several collaborators. In particular, 
I wish to thank Paolo Padovani, Bruno Arsioli and Matteo Perri, with whom I have been working on several of the topics reported in this paper.
I acknowledge the use of archival data and software tools from the ASDC, a facility managed by the Italian Space Agency (ASI). Part of this work is based on archival 
data from the NASA/IPAC Extragalactic Database (NED) and from the Astrophysics Data System (ADS). 
A large amount of archival data for the blazar 3C454.3 has been obtained from 
the OVRO (http://www.astro.caltech.edu/ovroblazars/), SMARTS (http://www.astro.yale.edu/smarts/glast/), Mets\"ahovi (http://www.metsahovi.fi/quasar/) 
and Catalina (http://crts.caltech.edu/) on-line services, which provide intensity measurements from long-term blazar monitoring programs based on an open data policy.

I thank the anonymous referee for his/her useful comments and suggestions that helped me improving the paper.

\newpage 

\section*{References}

\bibliography{Swift10years}

\end{document}